\documentclass[pra,aps,showpacs,twocolumn,floats,floatfix,superscriptaddress]{revtex4}
\usepackage{amsfonts}
\usepackage{amsmath}
\usepackage{amssymb}
\usepackage{graphicx}
\usepackage{txfonts}

\begin{document}

\title{Near Heisenberg limited parameter estimation precision by a dipolar
Bose gas reservoir engineering}
\author{Qing-Shou Tan}
\affiliation{College of Physics and Electronic Engineering, Hainan Normal University,
Haikou 571158, China}
\author{Ji-Bing Yuan}
\affiliation{Department of Physics and Electronic Information Science, Hengyang Normal
University, Hengyang 421002, China}
\author{Guang-Ri Jin}\email{grjin@bjtu.edu.cn}
\affiliation{Department of Physics, Beijing Jiaotong University, Beijing 100044, China}
\author{ Le-Man Kuang}\email{lmkuang@hunnu.edu.cn}
\affiliation{Key Laboratory of Low-Dimensional Quantum Structures and Quantum Control of
Ministry of Education, and Department of Physics, Hunan Normal University,
Changsha 410081, China}

\date{\today }

\begin{abstract}
We propose a scheme to obtain the Heisenberg limited parameter estimation precision by immersing atoms in a thermally equilibrated
quasi-one-dimensional dipolar Bose-Einstein condensate reservoir. We show
that the collisions between the dipolar atoms and the immersed atoms can
result in a controllable nonlinear interaction through tuning the relative
strength and the sign of the dipolar and contact interaction. We find that
the repulsive dipolar interaction reservoir is preferential  for the spin
squeezing and the appearance of an entangled non-Gaussian state. As an
useful resource for quantum metrology, we also show that the non-Gaussian
state results in the phase estimation precision in the Heisenberg scaling,
outperforming that of the spin-squeezed state.
\end{abstract}

\maketitle




\section{introduction}

One of main goals of quantum metrology is to achieve parameter (or phase) estimation precision beyond the shot-noise limit~\cite%
{Caves1981,Yurke1985,Holland1993,Giovannetti2011,Dorner2009,Sanders1995,Ma2011,Ma2011B,Humphreys2013,Dowling2008,Lvovsky2009}.
Atomic spin squeezed states (SSSs) play an important role in quantum phase estimation and
have been widely studied ~\cite{Kitagawa1993,Wineland1994,Momer,Sorensen2001,Poulsen2001,Jin2009,Jin2010,Guehne2009,Wang2003,Liu2011} in the past few decades ever since the pioneer work of Kitagawa and Ueda~\cite{Kitagawa1993}, who showed that the SSSs can be dynamically generated from the so-called one-axis twisting interaction (OAT) among spin-$1/2$ particles~\cite{bou,kas}. As useful quantum resource, the SSSs have been proposed to achieve such a sub-shot-noise limited phase sensitivity~\cite{Wineland1994,Momer,Sorensen2001,Poulsen2001,Jin2009,Jin2010,Guehne2009,Wang2003,Liu2011}. Recently, Stroble \textit{et al.}~\cite{strobel} experimentally demonstrated
that the OAT can also generate entangled non-Gaussian states (ENGSs), which can outperform the spin-squeezed state.
The OAT interaction have been proposed and demonstrated using  ion traps~\cite{Momer}, Rydeberg atoms~\cite{macri}, nitrogen-vacancy centers~\cite{Bennett} and atomic Bose-Einstein condensates (BECs).

BECs due to their unique coherence properties and the controllable nonlinearity~%
\cite{bou,kas}, have attracted much attention for quantum metrology.
Experimental
realizations of the OAT model has been proposed  and demonstrated
 through Feshbach resonances \cite{gros} or spatially separating
the components of BECs \cite{rie}.
Besides, the atomic BECs also often as the
reservoirs suitable for engineering is considered widely~\cite{Palzer,Will,Cirone,Spethmann,Scelle,Zipkes,Schmid,Balewski,Recati,Ng,bargill,Bru }. For instance, one can drastically enhance the OAT interaction by placing a two-state condensate in a
  completely different special BEC reservoir~\cite{bargill}.

So far, the studies of bosonic atoms for metrology are mainly focused on $s$%
-wave contact interaction. However, for ultracold atoms there also exists
long-rang magnetic dipole-dipole interaction (MDDI)~\cite%
{yi2000,Goral2002,yi,yi2007,lu2}. In experiments, dipolar BECs have been
realized for atoms with large magnetic dipole moments~\cite{Lu,Aikawa}.
Furthermore, both the sign and the strength of the effective dipolar
interaction can be tuned via a fast rotating orienting field \cite%
{Giovanazzi,Griesmaier2006,Lahaye}. Very recently, Yuan \textit{et al.} have
used the quasi-2D dipolar BEC as reservoir engineering to study the
non-Markovian dynamics of an impurity atom \cite{yuan}.
Therefore, the effects of the MDDI should be considered in the realizations of the OAT model
based on dipolar BEC.

In this paper, we realize  the OAT model induced by the reservoir dephasing noise,
which has been widely viewed as one of the main obstacles for quantum metrology.
We consider the dynamics of two-mode BEC consisting of $N $
atoms coupled to a one-dimensional (1D) dipolar Bose gas reservoir. It show
that, the collisions interaction between the dipolar BEC reservoir and the
immersed atoms can be described by a  spin-boson
model. Through calculating the SS $\xi_R$, and the quantum Fisher information
(QFI) $F_Q$, we find that the dephasing noise can produce SSSs and ENGSs. And the degree of the SS and entanglement both depends on
the relative strength and sign of the dipolar and contact interaction. In
other words, the repulsive dipolar interaction reservoir can induce better
SS and ENGSs. Compared with spin squeezed states, ENGSs can last for a very
long time under the dephasing noise. It can monotonically increase in the
regimes without SS ($\xi_R>1$), next successively undergoes metastable
entangled states and entanglement suddenly increase, corresponding to $F_Q
\simeq N^2/2$ and $F_Q\sim N^2$, respectively. According to Cram\'er-Rao theorem, $\Delta
\theta \geq 1/\sqrt{F_Q}$, we know that $F_Q>N$ means that the states are
entangled and useful for sub-shot-noise-limited phase-estimation precision;
and $F_Q = N^2$ is the maximal entangled states, corresponding to the
Heisenberg limit. This confirms that the phase estimation sensitivity can
 approach to Heisenberg limit, when using ENGSs for
metrology.

The paper is organized as follows. In Sec.~II, we give the model of immersed
atoms interacting with the thermally equilibrated quasi-1D dipolar BEC
reservoir. In Sec.~III, we study the dynamics evolution of the atoms due to
the dephasing noise. The SS and entanglement dynamical behaviors are
discussed in Sec.~IV and Sec.~V. Finally, we draw our conclusion in Sec.~VI.

\begin{figure}[htp]
\includegraphics[bb=30 100 570 500, width=3.2 in] {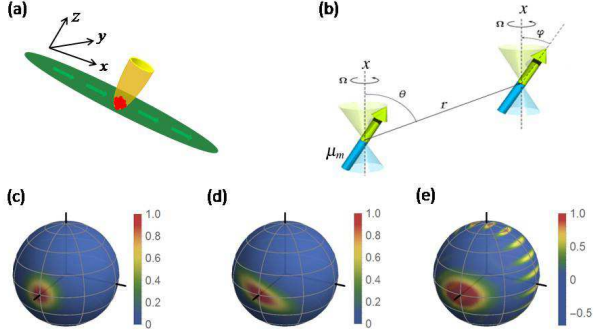}
\caption{ Schematic diagrams of (a) $N$ two-level atoms (red)
immersed in a quasi-1D dipolar gas (green) and (b) the tuning of the dipole-dipole
interaction via a fast rotating orienting field. (d)-(e) show the Wigner
function of the initial coherent spin state, the spin squeezed state, and
entangled non-Gaussian state (similar to the spin cat state) for the two-level atom system. The negative
values of Wigner function correspond to the quantum states.}
\end{figure}

\section{ formulation}

We consider a system of $N$ two states $^{87} \mathrm{Rb}$ atoms with up $%
|\uparrow \rangle \equiv |F=2,m_F=-1\rangle$ and down states $|\downarrow
\rangle \equiv |F=1,m_F=1\rangle$ immersed in a quasi-1D dipolar gas
reservoir. And the system atoms are confined in a harmonic trap that is independent of the
 internal states [see Fig.1(a)]. In general, the interaction between the system atoms and the
reservoir is described by the Hamiltonian
\begin{equation}
{H}= {H}_A+{H}_B+ {H}_{AB},
\end{equation}
where $H_A$ is the two-state atomic Hamiltonian, $H_B$ is the dipolar gas
reservoir Hamiltonian, and $H_{AB}$ describes their interaction.

\subsection{Two level atom system}

The spin Hamiltonian provides the most intuitive description in the internal
case of two-mode trapped atomic BEC
\begin{equation}
{H}_{A}=\lambda {J}_z+\chi {J}_{z}^{2}.
\end{equation}%
Here, we define the pseudo-spin operator ${\mathbf{J}}\equiv ({J}_{x}, {J}%
_{y}, {J}_{z})$ based on space orbitals as ${\mathbf{J}}=(a^{\dagger}_{%
\uparrow}, a^{\dagger}_{\downarrow}){\mathbf{\sigma}}(a_{\uparrow},
a_{\downarrow})^{T}/2$, where $\mathbf{\sigma}$ is the Pauli matrices, and $%
\hat{a}_{\uparrow}(\hat{a}_{\downarrow})$ denotes the annihilation operators
of the atom states. The energy difference of the two states $\lambda$ and
nonlinearity $\chi $ depend on the mean filed wave-function of the two
modes. We assume that the two modes have the same spatial orbital
\begin{equation}  \label{wf}
\Phi _{A}=(\pi \ell_{A}^{2})^{-3/4}e^{-(x^{2}+y^{2}+z^{2})/(2\ell_{A}^{2})},
\end{equation}
where $\ell_A=\sqrt{\hbar/(m_A\omega_A)}$ with $\omega_A$ being the trap
frequency, and $m_A$ being the mass of atom. Therefore, $\chi
=(g_{11}+g_{22}-2g_{12})/[2(2\pi )^{3/2}\ell_{A}^{3}]$
with coupling constants $g_{ij}=4\pi \hbar ^{2}a_{ij}/m_{A}$ and $a_{ij}$
being $s$-wave scattering length. For $^{87}\mathrm{Rb}$ and the chosen
hyperfine states, $a_{ij}$ are almost equal: $%
a_{11}=100.44a_{0},a_{22}=95.47a_{0}$ and $a_{12}=97.7a_{0}$, with $a_{0}$
being the Bohr radius. Then, the nonlinearity $\chi $ is close to zero.
We point out that  $\chi $ is tunable via Feshbach resonance,
but the price of these methods is significantly increased atom losses~\cite{bargill}.
Below, we choose $\chi =0,$ and apply a dipolar gas reservoir to induce a
stronger nonlinearity interaction.

\subsection{Bogoliubov modes of quasi-1D dipolar gas reservoir}

In second-quantized form, the many-body Hamiltonian of the 1D dipolar BEC is
\begin{eqnarray}
{H}_{B} &=&\int dx\hat{\Psi}_{B}^{\dagger }(x)\hat{h}\hat{\Psi}_{B}(x)
\notag  \label{Hami1} \\
&&+\frac{1}{2}\int dxdx^{\prime }\hat{\Psi}_{B}^{\dagger }(x)\hat{\Psi}%
_{B}^{\dagger }(x^{\prime })V(x-x^{\prime })\hat{\Psi}_{B}(x^{\prime })\hat{%
\Psi}_{B}(x),
\end{eqnarray}%
where $\hat{\Psi}_{B}(x)$ is the field operator and $\hat{h}=-\frac{\hbar
^{2}\partial ^{2}}{2m_{B}\partial x^{2}}$ is the single-particle Hamiltonian
with $m_{B}$ being the mass of the atom. Here, we have assumed that the
dipolar BEC to be confined in a cylindrically symmetric trap with a
transverse trapping frequency $\omega_{\perp}$ and negligible longitudinal
confinement  $\omega_x$ along the $x$ direction, i.e., $\omega_{\perp}/\omega_x \gg 1$.
In three dimensions, the two-body
interaction is
\begin{equation}
V^{3D}(\mathbf{r})=g_{B}\delta (\mathbf{r})+\frac{3c_{d}}{4\pi }\frac{1-3(%
\mathbf{\hat{\mu}_m\cdot \hat{r}})^{2}}{r^{3}},
\end{equation}
where the contact interaction strength is $g_{B}=4\pi \hbar ^{2}a_{B}/m_{B}$
with $a_{B}$ being the $s$-wave scattering length; the dipolar interaction
strength is $c_{d}=4\pi \hbar ^{2}a_{dd}/m_{B}$, where $a_{dd}=\mu _{0}\mu
_{m}^{2}m_{B}/(12\pi \hbar ^{2})$ is a length scale characterizing the MDDI
with $\mu _{0}$ the vacuum permeability, $\mu _{m}$ the magnetic dipole
moment; here $\hat{\mathbf{r}}=\mathbf{r}/r$ is a unit vector.

To obtain the effective 1D interaction potential, $V(x-x^{\prime })$, in
Hamiltonian~(\ref{Hami1}). We assume that the transverse wave function of
all the reservoir atoms is
\begin{equation}
\Psi _{\perp }(y,z)=(\pi
\ell_{B}^{2})^{-1/2}e^{-(y^{2}+z^{2})/(2\ell_{B}^{2})},
\end{equation}
with $\ell_{B}\equiv\sqrt{\hbar/(m_B \omega_{\perp})}$ being the width of
the Gaussian function. By integrating out the $y$ and $z$ variables, we can
obtain the Fourier transform of the 1D interaction potential (as shown in Appendix A)
\begin{eqnarray}
\tilde{V}_{1D}(k)=\frac{g_{B}}{2\pi \ell_{B}^{2}}\left[ 1- \epsilon_{dd}
\tilde{\nu}(k)\right]
\end{eqnarray}
with $\epsilon_{dd} \equiv c_{d}/g_{B}=a_{dd}/a_{B}$, where
\begin{eqnarray}
\tilde{\nu}(k)=1-\frac{3}{2}k^{2}\ell_{B}^{2}\exp\left[\frac{%
k^{2}\ell_{B}^{2}}{2} \right]\Gamma \left( 0,\frac{k^{2}\ell_{B}^{2}}{2}%
\right),
\end{eqnarray}
with  $\Gamma(0,x)$ being the incomplete Gamma function.

To proceed, in the degenerate regime, the bosonic field can be decomposed as
\begin{equation}  \label{psib}
\hat{\Psi}_{B}(\mathbf{r})=\Psi _{\perp }(y,z)\left[ \sqrt{n_{0}}+\frac{1}{%
\sqrt{L}}\sum_{k}\left(u_k \hat{b}_{k}e^{ikx}-v_k\hat{b}^{%
\dagger}_{k}e^{-ikx}\right)\right],
\end{equation}
with $n_{0}$ being the condensate linear density, $L$ the length of the
reservoir, $\hat{b}_{k}$ ($\hat{b}^{\dagger}_{k}$) the annihilation
(creation) operators of the Bogoliubov modes with momentum $k$. And its
Bogoliubov modes are
\begin{eqnarray}
u_k&=&1/2\left(\sqrt{\varepsilon_k/E_k}+\sqrt{E_k/\varepsilon_k}\right),\nonumber\\
v_k&=&1/2\left(\sqrt{\varepsilon_k/E_k}-\sqrt{E_k/\varepsilon_k}\right),
\end{eqnarray}
with $E_{k}=\hbar
^{2}k^{2}/(2m_{B})$ being the free-particle energy. Where the excitation
energy is~\cite{Cirone}
\begin{eqnarray}
\varepsilon _{k} &=&\sqrt{E_{k}^{2}+2n_{0}E_{k}\tilde{\nu}(k)}  \notag \\
&=&\frac{1}{2}{\hbar \omega _{\perp}}\sqrt{(k \ell_{B})^{4}+ \eta
(k\ell_{B})^{2}\left[ 1-\epsilon _{dd}\tilde{\nu}_{D}(k) \right] },
\end{eqnarray}
with dimensionless parameters $\eta=8n_{0}a_{B}$. Hence, the Hamiltonian for
the collective excitations is
\begin{equation}
{H}_{B}^{\prime }=\sum_{k\neq 0}\varepsilon _{k}b_{k}^{\dag }b_{k}.
\end{equation}
The sum over Bogoliubov modes exclude the zero mode and will act as the
reservoir under our model.

\subsection{Interaction Hamiltonian}

We assume that the reservoir atoms are coupled with the up state $%
|\uparrow \rangle $ of the system atoms via a Raman transition~\cite{Cirone,yuan}
\begin{eqnarray}
{H}_{AB} &=&g_{AB} \hat{a}_{\uparrow }^{\dag }\hat{a}_{\uparrow } \int d\mathbf{r} |{\Phi%
}_{A}(\mathbf{r})|^2 \hat{\Psi}_{B}^{\dag }(\mathbf{r})\hat{\Psi}_{B}(\mathbf{r}),
\end{eqnarray}
where $g_{AB}=2\pi \hbar ^{2}a_{AB}/m_{AB}$ with the atoms and reservoir
scattering length $a_{AB}$ and reduced mass $%
m_{AB}=m_{A}m_{B}/(m_{A}+m_{B}). $
By substituting Eqs.~(\ref{wf}) and~(\ref{psib}) into the above interaction Hamiltonian and omitting the square terms about
$\hat{b}_k$ and $\hat{b}^{\dagger}_k$,
we have
\begin{eqnarray}
{H}_{AB}&\simeq & \delta _{\uparrow }\hat{a}_{\uparrow }^{\dag }\hat{a}_{\uparrow }+%
\hat{a}_{\uparrow }^{\dag }\hat{a}_{\uparrow } \sum_{k} g_{k}\left(\hat{b}%
_{k}+\hat{b}_{k}^{\dag }\right),
\end{eqnarray}
 where
\begin{eqnarray}
\delta _{\uparrow } &=&{g_{AB}n_{0}}\int dydz\left\vert \Psi
_{B}(y,z)\right\vert ^{2}\left\vert \Phi _{A}(y,z)\right\vert ^{2}\int
dx\left\vert \Phi _{A}(x)\right\vert ^{2}  \notag \\
&=&\frac{2\hbar^2 a_{AB}n_{0}}{m_{AB}(\ell_{A}^{2}+\ell_{B}^{2})},
\end{eqnarray}
and
\begin{eqnarray}
g_{k} =\frac{2\hbar ^{2}a_{AB}}{m_{AB}(\ell_{A}^{2}+\ell_{B}^{2})} \sqrt{%
\frac{n_{0}E_{k}}{L\epsilon _{k}}}{\exp \left( -\frac{k^{2}\ell_{A}^{2}}{4}%
\right) }.
\end{eqnarray}

\section{System dynamical evolution}

In the interaction picture with respect to ${H}_{B}^{\prime }$, the total
Hamiltonian is
\begin{equation}
H_{I}(t)=(\lambda +\delta _{\uparrow })J_{z}+N_{\uparrow
}\sum_{k}g_{k}(b_{k}^{\dag }e^{i\omega _{k}t}+b_{k}e^{-i\omega
_{k}t})-i\Gamma _{\mathrm{loss}}N_{\uparrow },  \label{hi}
\end{equation}
which is  a non-Hermitian dephasing
spin-boson model with $N_{\uparrow }=\left( J_{z}+N/2\right)$ being the up state number operator.
Here the non-Hermitian term $\Gamma _{\mathrm{loss}}$ is phenomenally introduced to describe the one-boby particle loss rate, owing to
inelastic collisions between the system atoms and the noncondensed thermal atoms.
It results in the particles be kicked out from the system. Such a kind of   loss is
 a typical dissipation effect  and has been widely studied~\cite{strobel,pawlowski,hao,spehner, huang}.

 By using of Magnus expansion~\cite{bargill}, the time
evolution operator can be read as $U(t)=e^{-itH_{\mathrm{eff}}},$ where the
effective Hamiltonian is (see Appendix B for details)
\begin{equation}  \label{heff}
H_{\mathrm{eff}}=\lambda^{\prime }J_{z}+\Delta
(t)J_{z}^{2}+iJ_{z}\sum_{k}(\alpha _{k}b_{k}^{\dagger }-\alpha ^{\ast
}b_{k})-i\Gamma _{\mathrm{loss}}N_{\uparrow },
\end{equation}
with $\lambda^{\prime }\equiv\lambda+\delta_{\uparrow} -N\Delta (t)$ and $%
\alpha _{k}\equiv g_{k}(1-e^{i\omega _{k}t})/(t{\omega _{k}})$. From the
above equation, we can find that the collisional interaction between the
atoms and reservoir induce a nonlinear term $\propto J_{z}^{2}$,
corresponding to the OAT Hamiltonian, and the noise induced
nonlinear strength is~\cite{Breuer}
\begin{eqnarray}  \label{delta}
\Delta (t) = \frac{1}{t}\int_{0}^{\infty }d\omega J(\omega )\frac{\omega
t-\sin (\omega t)}{\omega ^{2}}.
\end{eqnarray}%
Here, the reservoir spectral density $J(\omega )$ defined as $J(\omega
)=\sum_{k\neq 0}\left\vert g_{k}\right\vert ^{2}\delta (\omega -\varepsilon
_{k}/\hbar )$. In the continuum limit, $L^{-1}\sum_{k}\rightarrow (2\pi
)^{-1}\int dk$, we have

\begin{eqnarray}
J(\omega ) &=&{\Theta}\hbar \omega _{\perp}^{3}\ell_{B}^{3}\int_{0}^{\infty
}dk\frac{ k^{2}e^{-k^{2}\ell_{A}^{2}/2}}{\varepsilon (k)}\delta \left(\omega
-\frac{\varepsilon(k)}{\hbar }\right)  \notag \\
&=&{\Theta}\hbar \omega _{\perp}^{3}\ell_{B}^{3}\sum_{i}\frac{f(k_{i}(\omega
))}{\omega } \left\vert \frac{d\varepsilon (k)}{dk}\right\vert
_{k=k_{i}(\omega )}^{-1},
\end{eqnarray}
where $f(k)\equiv k^{2}e^{-k^{2}\ell_{A}^{2}/2}$ with $k_{i}(\omega )$ being
the roots of the equation $\varepsilon (k)=\hbar \omega ,$ and the
dimensionless parameter is
\begin{equation}
{\Theta}=\frac{n_0 \ell_{B}^{3}a_{AB}^{2}(m_{A}+m_{B})^{2}}{\pi m_{A}^{2}
\left(\ell_{A}^{2}+\ell_{B}^{2}\right)^{2}}.
\end{equation}

Assuming that the initial state of the total system is given by
\begin{equation}
\rho _{T}(0)=\left\vert \Phi (0)\right\rangle _{A}\left\langle \Phi
(0)\right\vert \otimes \rho _{B},
\end{equation}
where $\left\vert \Phi (0)\right\rangle _{A}\equiv\frac{1}{2^{N/2}} \left(
\left\vert \uparrow \right\rangle +\left\vert \downarrow \right\rangle
\right)^{\otimes N}=\sum_{m}c_{m}(0)\left\vert j,m\right\rangle $ is CSS,
with the probability amplitudes $%
c_{m}=2^{-j}\left( C_{2j}^{j+m}\right) ^{1/2}$ and total spin $j=N/2$ for a
system consisting of $N$ condensated atoms. And the density matrix of
reservoir read as

\begin{equation}
\rho_{B}=\Pi _{k}[1-\exp (-\beta \omega _{k})]\exp (-\beta \omega
_{k}b_{k}^{\dag }b_{k}),
\end{equation}
with $\beta $ the inverse temperature. With the help of Eq.~(\ref{heff}), the time-evolution reduced matrix
elements of the atom system at any later time $t$ is found by tracing over the reservoir degrees of freedom

\begin{eqnarray}\label{rhot}
\rho _{jm,jn}^{A}(t)&=&e^{-it\lambda'(m-n)}e^{it\Delta
(t)(m^{2}-n^{2})}  \notag \\
&&\times e^{-\Gamma_{\mathrm{loss}} t(m+n+N)}e^{-t(m-n)^{2}\gamma (t)}\rho
_{jm,jn}^{A}(0),
\end{eqnarray}
with decoherence function
\begin{eqnarray}\label{gamt}
\gamma (t) &=&\frac{1}{t}\int_{0}^{\infty }d\omega J(\omega )\coth \left(
\frac{\hbar \omega }{2k_{B}T}\right) \frac{1-\cos (\omega t)}{\omega ^{2}},
\end{eqnarray}
which is a function of temperature $T$.

In Eq.~(\ref{rhot}), $\Delta (t)$ and $\gamma (t)$, respectively, correspond to the unitary and non-unitary evolution due to the effects of the reservoir.
Equations (\ref{delta}) and (\ref{gamt}) show that both $\Delta (t)$ and $\gamma (t)$ depend on the reservoir spectral density $J(\omega)$,
which can be controlled by tuning the MDDI of the dipolar Bose gas
reservoir ~\cite{Giovanazzi,Griesmaier2006,Lahaye}.
This is the main difference from Feshbach resonances method
whose control is on the system
atoms directly and
will strongly enhance three-body loss near
resonances regime beside the one-body loss.
Therefore, in our model the one-body particle loss due to the  inelastic collisions of the noncondensed atoms maybe the main loss mechanism that need  considering. When temperature is low enough
 the values of one-body loss rate $\Gamma _{\mathrm{loss}}$ will be very small, since  there is only a
 small number of thermal atoms with sufficient energy to knock atoms out of condensate~\cite{Dalfovo}.
For simplicity, hereafter we only
consider the case of $T\rightarrow 0$, and choose $\Gamma _{\mathrm{loss}}$ as  a free parameter.

\section{Spin squeezing and entangled non-Gaussian spin states}

\subsection{Spin squeezing parameter}

Now, a state is regarded as squeezed if the variance of one spin component
normal to the mean spin vector $\langle {\boldsymbol{J}}\rangle ={\mathrm{Tr}%
}\left[ \boldsymbol{J}\rho ^{A}(t)\right] $ is lower than the Heisenberg
limited value. The SS parameter defined by Wineland is~\cite{Wineland1994}
\begin{equation*}
\xi_R ^{2}=\frac{N(\Delta J_{\hat{n}_{\perp }})_{\mathrm{min}}^{2}}{|\langle
{\mathbf{J}}\rangle |^{2}},
\end{equation*}
where $(\Delta J_{\hat{n}_{\perp }})_{\mathrm{min}}$ represents the minimal
variance of the spin component perpendicular to the mean spin direction $%
\hat{r}_{0}\equiv \langle {\mathbf{J}}\rangle /|\langle {\mathbf{J}}\rangle
| $. Where the mean spin is $|\langle {\mathbf{J}}\rangle |=\sqrt{%
\left\langle J_{x}\right\rangle ^{2}+\left\langle J_{x}\right\rangle
^{2}+\left\langle J_{z}\right\rangle ^{2}}$. A state is spin squeezed if $%
\xi_R ^{2}<1$. In addition, the smaller $\xi_R ^{2}$ is, the stronger the
squeezing is. If $\Gamma_{\mathrm{loss}} =0$, the squeezing parameter can be
evaluated explicitly~
\begin{equation}\label{ss}
\xi_R ^{2}=\frac{4+(N-1)\left( \tilde{A}-\sqrt{\tilde{A}^{2}+\tilde{B}^{2}}
\right) }{4e^{-2t\gamma (t)}\left[ \cos (t\Delta (t))\right] ^{2N-2}},
\end{equation}
which does not depend on  $\lambda'$.
And the optimally squeezed direction is $\phi_{\rm opt}=[\pi+\tan^{-1}(\tilde{B}/\tilde{A})]$,
with
\begin{eqnarray}
\tilde{A} &=&1-\cos ^{N-2}[2(t\Delta (t))]\exp [-4t\gamma (t)],  \notag \\
\tilde{B} &=&-4\sin[t\Delta (t)]\cos ^{N-2}[t\Delta (t)]\exp [-t\gamma (t)].
\end{eqnarray}
When  considering the one-body losses, $\Gamma_{\mathrm{loss}} \neq 0,$ the form of SS has given
in Appendix A.

Equation~(\ref{ss}) indicates that the dephasing noise plays two different roles:
 On one hand, it can generate the SS by
creating the nonlinear interaction $\Delta (t)$;
on the other hand, it degrades the degree of SS via the
decoherence function $\gamma (t)$.

\begin{figure}[tph]
\center{\includegraphics[scale=0.7]{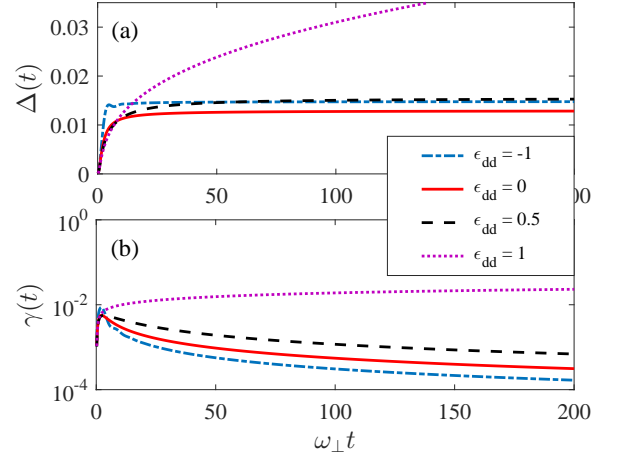}}
\caption{ Time dependence of $\Delta(t)$ and $\protect\gamma%
(t) $ for different $\epsilon_{dd}$. Here we choose $\Theta=1.5\times 10^{-2}$ and
$\eta=5$. Note that for repulsive MDDI  the values of noise-induced nonlinear interaction $\Delta(t)$ approach to  their steady values $\Delta(\infty)$, while decoherence function $\gamma(t)$ is decreasing with time.}
\end{figure}

\begin{figure}[tph]
\center{\includegraphics[scale=0.56]{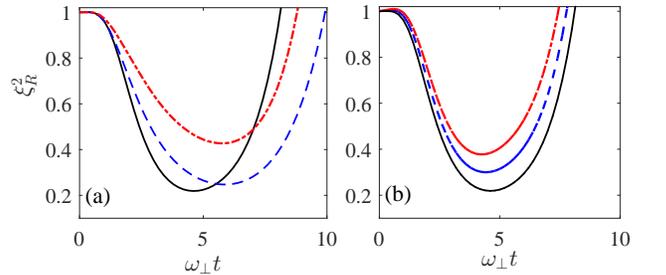}}
\caption{ Spin squeezing dynamics of two-mode BEC consisting
of $N=100$ atoms coupled to a 1D dipolar Bose gas reservoir. (a) Squeezing parameter for different
values of $\epsilon_{dd}$: from top to bottom $\protect\epsilon _{dd}=1,0,-1$.
Here $\Gamma_{\mathrm{loss}} =0$. (b) Squeezing parameter for different
values of loss parameters with $\protect\epsilon _{dd}=-1$, from top to
bottom $\Gamma_{\mathrm{loss}} =0.01\Delta(\infty),0.002\Delta(\infty)$, and
0.}
\end{figure}

\subsection{QFI and entanglement of non-Gaussian spin states}

To investigate the entanglement created by the dipolar BEC reservoir, we can
also introduce the QFI. In general, the states that are entangled and useful
for sub-shot-noise-limited parameter-estimation precision is identified by
the QFI criterion $F_Q>N$. The QFI $F_{Q}$ with respect to $\theta $,
acquired by an SU(2) rotation, can be described as~\cite{Ma2011B}
\begin{equation}
F_{Q}[\rho (\theta ,t),\hat{J}_{\vec{n}}]=\vec{n}\boldsymbol{C}\vec{n}^{T},
\end{equation}
where $\rho(\theta,t)=\exp(-i \theta J_{\vec{n}})\rho(t)\exp(i\theta J_(\vec{n}))$ with $\vec{n}$ being the optimal rotation direction, and the matrix element
for the symmetric matrix $\boldsymbol{C}$ is
\begin{equation}  \label{ckl}
C_{kl}=\sum_{i\neq j}\frac{(p_{i}-p_{j})^{2}}{p_{i}+p_{j}}[\left\langle
i\right\vert J_{k}\left\vert j\right\rangle \left\langle j\right\vert
J_{l}\left\vert i\right\rangle +\left\langle i\right\vert J_{l}\left\vert
j\right\rangle \left\langle j\right\vert J_{k}\left\vert i\right\rangle,
\end{equation}
with $p_{i}(|i\rangle )$ being the eigenvalues (eigenvectors) of $\rho
(\theta ,t)$.

For simplicity, we first consider the case of small $N$, e.g., $N=2$. When
setting $\lambda'=0$ and neglecting the
particle loss, e.g., $\Gamma_{\mathrm{loss}} =0$, the QFI can be calculated
analytically
\begin{equation}
F_{Q}[\rho (\theta ,t),\hat{J}_{n}]=\max [C_{xx},C_{\perp }],
\end{equation}
with
\begin{equation}
C_{xx}=\frac{4\sinh ^{2}[2\gamma (t)t]+16e^{2\gamma (t)t}}{1+3e^{4\gamma
(t)t}}\left[ 1-\frac{16\cos ^{2}[\Delta (t)t]}{e^{-6\gamma
(t)t}[1-e^{4\gamma (t)t}]^{2}+16}\right]
\end{equation}
in $x$-axis direction, and
\begin{equation}
C_{\perp }=\frac{C_{yy}+C_{zz}+\sqrt{(C_{yy}+C_{zz})^{2}+4C_{yz}^{2}}}{2}
\end{equation}
in $yz$ plane, which can also be obtained by using of Eq.~(\ref{ckl}) (see
Appendix B).

The maximal QFI can be found in $x$ axis direction
\begin{equation}  \label{fq}
F_{Q}^{\mathrm{max}}=\frac{4\sinh ^{2}[2\gamma (t)t_{\mathrm{opt}
}]+16e^{2\gamma (t)t_{\mathrm{opt}}}}{1+3e^{4\gamma (t)t_{\mathrm{opt}}}}
\end{equation}
when choosing optimal interrogation time $t_{\mathrm{opt}}=\pi /[2\Delta
(t)] $. Equation (\ref{fq}) reveals that the values of $\gamma(t)$ is
directly related to the QFI. One finds $F_{Q}^{\mathrm{max} }\rightarrow
N^{2}$ (the Heisenberg limit) if $\gamma (t)\rightarrow 0$. Fortunately, Fig.~2(b) indicates
that  the nearly neglectable  $\gamma (t)$ can be obtained  when the dipolar Bose gas reservoir with repulsive MDDI.
Therefore, the main limitation of Heisenberg scaling is one-body loss   mechanism in our scheme.
Next, we consider large $N$ and the case of $\Gamma_{\mathrm{loss}} \neq 0$
numerically.

\section{Results and discussion }

As a concrete example, we consider a BEC reservoir of $^{162}\mathrm{Dy}$
atoms, for which we have $\mu_m=9.9\mu _{B}$ \ and $a_{B}=112a_{0}$ with $%
\mu_B$ the Bohr magneton. This means that $a_{dd}\simeq 131a_{0}\ $ and
dipolar interaction is mainly attractive since $\epsilon_{dd}>1,$ then the
attraction is stronger than the short-range repulsion. However, not only the
contact interaction strength can be tunable via Feshbach resonance, but also
both the sign and the strength of the effective dipolar interaction can be
tuned by the use of a fast rotating orienting field. Later, we will consider
the values of $\epsilon _{dd}\in \lbrack -1,1]$, which is repulsive
(attractive) MDDI for $\epsilon_{dd}<0$ ($\epsilon_{dd}>0$).


\begin{figure}[tph]
\center{\includegraphics[scale=0.47]{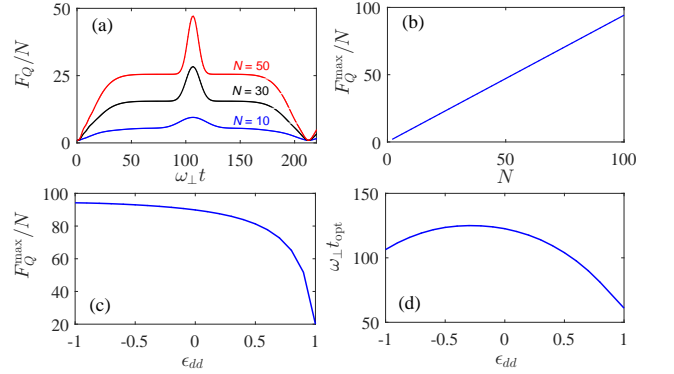}}
\caption{ (a) Time dependence of QFI amplification rate with
repulsive interaction $\protect\epsilon_{dd}=-1$. We show three cases with $%
N=10, 30$ and $50$. (b) The maximal QFI amplification rate $F^{\mathrm{max}}_Q/N$
as a function of atom number $N$ with $\protect\epsilon_{dd}=-1$. (c) QFI
amplification rate and (d) the corresponding optimal evolution time $\protect%
\omega_{\perp} t_{\mathrm{opt}}$ with respect to $\protect\epsilon_{dd}$
when $N=100$. Here $\Gamma_{\mathrm{loss}}=0$.}
\end{figure}

Numerically, it is convenient to introduce the dimensionless units: $\hbar
\omega _{\perp }$ for energy, $\omega _{\perp }^{-1}$ for time, and $\ell
_{B}=[\hbar /(m\omega _{\perp })]^{1/2}$ for length. To obtain the values of
dimensionless parameter $\eta$ and $\Theta$, we assume a quasi-1D trap  with
$\omega_x=2\pi \times 20\mathrm{Hz}$ and $\omega _{\perp }=2\pi \times 10^{3}\mathrm{Hz}$;
and the corresponding
harmonic oscillator width is $\ell _{B}=\ell_{A}\simeq 2.5\times 10^{-7}%
\mathrm{m}$. We assume that the linear density of the quasi-1D condensate is
$n_0=10^{8}\mathrm{m^{-1}}$, and the $s$-wave scattering length between Rb
and Dy atoms is $a_{AB}\sim \mathrm{5nm }$~\cite{yuan}. Then, we shall take $%
\eta=5 $ and $\Theta=1.5\times 10^{-2}$ in the results presented below.

Since both the SS and QFI depend on $\gamma (t)$ and $\Delta (t),$ let us
first investigate the time dependence of dephasing factor. From Fig.~2, we
can clearly see that the squeezing rate $\Delta(t)$ is nearly constant, $\Delta(\infty)$.
Whereas $\gamma(t)$ is decreasing with time, what is more we can get
very small $\gamma(t)$  (e.g.$10^{-3}-10^{-4}$) when $\epsilon_{dd}<0$.
 Comparing $\Delta(t)$ and $\gamma(t)$, we can also find
that the values of $\Delta(t)$ (squeezing rate) is larger than $\gamma(t)$
(dispersive rate) for $\epsilon_{dd}<0$, which means that we can obtain
strong squeezing and large QFI by the reservoir's engineering with repulsive
MDDI.

In Fig.~3, we plot the SS $\xi^2_{R}$ dynamics of two-mode BEC consisting $N$
atoms coupled to a 1D dipolar Bose gas reservoir for various $\epsilon_{dd}$
values. As is shown in Fig.~3(a), the optimal squeezing can be achieved
within short time scale, and after a transient time, it is lost ($\xi_R^2>1$%
) and then ENGSs are produced. However, for repulsive-dipolar-interaction
reservoir, we can obtain stronger SS, due to their
smaller dispersive rate $\gamma(t)$. In Fig.~3(b), we further plot the time
evolution of $\xi^2_R$ for various atom loss rates. It indicates that the
squeezing degree degraded with the increasing of the atom loss rate.

Figure 4 illustrates the QFI amplification rates with respect to the initial
state (CSS) ${F}_Q/N$. In Fig.~4(a), we present the time dependence of QFI
amplification rates with repulsive interaction. We see that differ from the
case of SS, the amplified QFI can last for a very long time, which means
that the ENGSs can be produced and achieve the maximal even in the regimes
without squeezing. The optimal QFI first monotonically increasing and then
reach a metastable $\sim N/2$ in the $yz$ plane. Subsequently, the QFI
suddenly increasing in the $x$-axis direction at the optimal interrogation
time $t_{\mathrm{opt}}$. As shown in Fig.~4(b), the maximal amplification
rate $F^{\mathrm{max}}_Q/N$ is proportional to the atom number $N$, and the
scale factor is $\sim N$, which is the Heisenberg scale. What is more,
compared with the attractive MDDI reservoir the repulsive interaction
can induce larger QFI, this result is presented in Fig.~4(c).

Comparison of  Figs.~4(a) and 5(a), we can see that for not too
large atom loss rates $\Gamma_{\mathrm{loss}}$ the optimal evolution time
is $t_{\mathrm{opt}}=\pi /[2\Delta(t)] $, which is the same as the case of $%
N=2$.
\begin{figure}[tph]
\center{\includegraphics[scale=0.56]{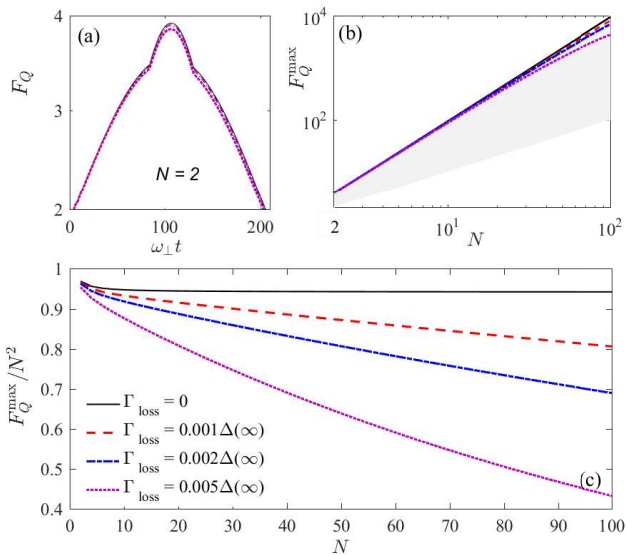}}
\caption{ (a) QFI vs time $\omega_{\perp}t$ for various $\Gamma_{\mathrm{loss}}$ with $N=2$.
  The solid line (black) corresponds to the analytic solution given in Eqs.~(30-33).
(b) The maximal QFI as a function of atom number $N$ for various $\Gamma_{\mathrm{loss}}$.
 The shaded area indicates the regime between shot-noise limit and Heisenberg limit.
 (c) The rates of the Heisenberg limit $F^{\mathrm{max}%
}_Q/N^2 $ with respect to atom number $N$ for different values of loss
parameters $\Gamma_{\mathrm{loss}}$. Other parameters are $\protect\epsilon_{dd}=-1$,
 from top to
bottom $\Gamma_{\mathrm{loss}}=0, 0.001\Delta(\infty), 0.002\Delta(\infty),$ and $0.005\Delta(\infty)$. }
\end{figure}

Figures~5(b) and (c) illustrate the QFI
with respect to atom number $N$ for different values of loss parameters.
As shown in Fig.~5(b), under the values of $\Gamma_{\mathrm{loss}}$ we considered, we can obtain
near-Heisenberg scaling.
Figures~5(c) plots the  rates of the Heisenberg limit $F^{\mathrm{max}%
}_Q/N^2 $ with respect to atom number $N$ for different values of loss
parameters $\Gamma_{\mathrm{loss}}$. It indicates that for not too large $\Gamma_{\mathrm{loss}}$, we can obtain the Heisenberg scaling only with a prefactor.
We can also see, when increasing the loss rate the values of $F^{\mathrm{max}%
}_Q/N^2 $  degrades with the increasing of number of atoms $N$, since
the collective dissipate rare $N \Gamma_{\mathrm{loss}}$ depends on $N$.
Fortunately,  for the low temperature limit we considered  the values of $\Gamma_{\mathrm{loss}}$ is not too large, hence we can
still obtain the robust sub-shot-noise-limited phase sensitivity.
\begin{figure}[tph]
\center{\includegraphics[scale=0.55]{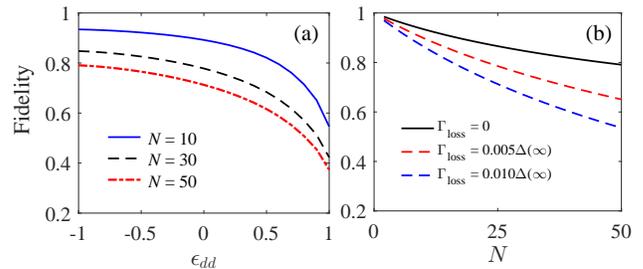}}
\caption{(a) The fidelity between our optimal ENGSs and spin cat
state  with respect to $\protect\epsilon_{dd}$ for different values of atom number $N$. Here we choose $\Gamma_{\rm loss}=0$.
(b) The fidelity as a function of atom number $N$ for various $\Gamma_{\rm loss}$ with $\protect\epsilon_{dd}=-1$.}
\end{figure}

To demonstrate the near Heisenberg-limited sensitivity with the ENGSs
realized by our model, we calculate the fidelity between the optimal ENGSs
and spin cat states~\cite{Momer}, which are the maximal entangled states and have the
Heisenberg-limited sensitivity for metrology. It is given by
\begin{equation}
|\Psi \rangle _{\mathrm{cat}}=\frac{1}{\sqrt{2}}\left( \left\vert \frac{\pi
}{2},0\right\rangle -e^{-i\frac{\pi }{2}(N+1)}\left\vert \frac{\pi }{2},\pi
\right\rangle \right) ,
\end{equation}%
where $|\theta _{0},\phi _{0}\rangle \equiv e^{i\theta _{0}(J_{x}\sin \phi
_{0}-J_{y}\cos \phi _{0})|j,j\rangle }$ is the CSS. With the definition of
fidelity, we have
\begin{equation}
\mathcal{F}_{\varrho }=\mathrm{tr}\sqrt{\varrho _{\mathrm{cat}}^{1/2}\rho
^{A}(t_{\rm opt})\varrho _{\mathrm{cat}}^{1/2}},
\end{equation}%
where $\varrho _{\mathrm{cat}}=|\Psi \rangle _{\mathrm{cat}}\langle \Psi |$.

Form Fig.~6, we can find that the fidelity depends on both the $\epsilon
_{dd}$ and particle number $N$. The maximal values occurs at $\epsilon
_{dd}=-1$.  Because of the dissipative rate $\gamma (t)$, the fidelity
decrease with the increase of $N$.  As shown in
Fig.~6(a), when $\Gamma_{\rm loss}=0$ we get $\mathcal{F}_{\varrho }\simeq 0.94,0.85$, and $0.80$ for $%
N=10,30 $, and $50$.
And the effect of $\Gamma_{\rm loss}$ on fidelity
also is shown in Fig.~6(b).

\section{conclusion}

In summary, we have realized  the SSSs and ENGSs by immersing atoms in a thermally equilibrated
quasi-1D dipolar BEC reservoir.
 It has demonstrated that the
repulsive-dipolar-interaction reservoir can induce better SS and entanglement.
 We have shown that owing to the dephasing
noise, even in the regimes without SS the ENGSs can successively undergo
highly metastable entangled states and entanglement suddenly increase.
To explain the highly sensitivity for metrology,  we calculated the fidelity between the optimal ENGSs
and spin cat states, and found that
the optimal ENGS is similar to the spin cat state.
It has confirmed that by the use of ENGSs for metrology, the phase estimation
sensitivity can surpass that by SS and even approach to Heisenberg limit for  neglectable atom loss rates. The
effect of the atom loss rate as a free parameter has also been considered.

Finally,  we give two remarks on the above obtained results:
First, these results we have obtained in this paper are based on the
negligible spatial evolution of the immersed condensate
wave functions. Usually this assumption is enough to capture the basic processes and physics, and
detailed consideration of the  impact of spatial dynamics  can be investigated by adopting  the
multi-configurational time-dependent Hartree for bosons method~\cite{stre2006}.
Second, the scheme we  proposed in this work can also suit the case that the system atoms are
weak or no interaction two-level impurity atoms which are not condensate.

\begin{acknowledgments}
This work was supported by the NSFC under Grant No. 11547159. G.R.J.
acknowledges support from the Major Research Plan of the NSFC (Grant No.
91636108).
\end{acknowledgments}

\appendix
\begin{widetext}

\section{Derivation of Eq.~(7)}
Here, we present a detailed derivation of the Fourier transform of the effective 1D interaction potential.
It can be
obtained by integrating out the $y$ and $z$ variables as

\begin{eqnarray}
\tilde{V}_{1D}(k) &=&\frac{1}{2\pi }\int dydz\left\vert \Psi _{\perp
}(y,z)\right\vert ^{2}F_{yz}^{-1}\left[ F_{yz}\left[ \left\vert \Psi _{\perp
}(y,z)\right\vert ^{2}\right] \tilde{V}(k)\right]  \notag \\
&=&\frac{1}{2\pi }\int dydz\left\vert \Psi _{\perp }(y,z)\right\vert
^{2}F_{yz}^{-1}\left[ \frac{1}{2\pi }%
e^{-(k_{y}^{2}l_{B}^{2}+k_{z}^{2}l_{B}^{2})/4}\left[ g_{B}-c_{d}\left( 1-3(%
\mathbf{\hat{\mu}}_{m}\mathbf{\cdot \hat{e}}_{k})^{2}\right) \right] \right].
\end{eqnarray}%
When assuming the dipole moments lie on the $xz$ plane forming an angle $%
\varphi $ to $x$ axis, i.e.,
\begin{equation}
\mathbf{\hat{\mu}}_{m}=(\cos \varphi ,0,\sin \varphi ).
\end{equation}%
Then, we have
\begin{eqnarray}
\tilde{V}_{1D}(k) &=&\frac{1}{\left( 2\pi \right) ^{2}}\int d\phi \int
dk_{\perp} k_{\perp} e^{-(k_{\perp} ^{2}l_{B}^{2})/2}\left[ g_{B}-c_{d}\left( 1-3\frac{%
(k\cos \varphi +k_{\perp} \cos \phi \sin \varphi )^{2}}{k^{2}+k_{\perp} ^{2}}%
\right) \right]  \notag \\
&=&\frac{g_{B}}{2\pi l_{B}^{2}}-\frac{c_{d}}{2\pi l_{B}^{2}}\left( 1-\frac{3%
}{2}\sin ^{2}\varphi \right) \left[ 1-\frac{3}{2}k^{2}l_{B}^{2}\exp \left(
\frac{k^{2}l_{B}^{2}}{2}\right) \Gamma \left( 0,\frac{k^{2}l_{B}^{2}}{2}%
\right) \right]  \notag \\
&=&\frac{g_{B}}{2\pi l_{B}^{2}}-\frac{\tilde{c}_{d}}{2\pi l_{B}^{2}}\left[ 1-%
\frac{3}{2}k^{2}l_{B}^{2}\exp \left( \frac{k^{2}l_{B}^{2}}{2}\right) \Gamma
\left( 0,\frac{k^{2}l_{B}^{2}}{2}\right) \right] \notag \\
&=&\frac{g_{B}}{2\pi l_{B}^{2}}\left\{1-\tilde{\epsilon}_{dd}\left[ 1-%
\frac{3}{2}k^{2}l_{B}^{2}\exp \left( \frac{k^{2}l_{B}^{2}}{2}\right) \Gamma
\left( 0,\frac{k^{2}l_{B}^{2}}{2}\right) \right]\right\},
\end{eqnarray}%
where $\tilde{\epsilon}_{dd}=\tilde{c}_d/g_B$ with
$k_{\perp}=\sqrt{k_y^2+k_z^2}$ and $\tilde{c}_{d}=c_{d}\left( 1-\frac{3}{2}\sin ^{2}\varphi \right)$.
Clearly, the effective 1D dipolar interaction vanishes  at the magical angle
  $\alpha_{m}=54.74^{\circ}$, and it
is attractive (repulsive)  for $\alpha<\alpha_{m}(\alpha>\alpha_{m})$.
In the main text, we have dropped the tilde on $\tilde{c}_{d}$ and $\tilde{\epsilon}_{dd}$, and will only consider the values of $\epsilon _{dd}\in [-1,1]$.

\section{Time evolution operator $U(t)$}

The time evolution operator can be obtained by using Magnus expansion
\begin{equation}\label{u1}
U(t)\equiv \mathrm{T}_{+}\exp \left[ -i\int_{0}^{t}H_{I}(t^{\prime
})dt^{\prime }\right] =\exp \left[ \sum_{n=1}^{\infty }\frac{(-i)^{n}}{n!}%
F_{n}(t)\right].
\end{equation}%
Note that only the below first two terms of the expansion are non-zero
\begin{eqnarray}
F_{1}(t) &=&\int_{0}^{t}H_{I}(t^{\prime })dt^{\prime }=\lambda tJ_{z}+\left(
J_{z}+\frac{N}{2}\right) t\sum_{k}\int_{0}^{t}\left( g_{k}b_{k}^{\dag
}e^{i\omega _{k}t^{\prime }}+g_{k}^{\ast }b_{k}e^{-i\omega _{k}t^{\prime
}}\right)   \notag \\
&=&\lambda tJ_{z}+\left( J_{z}+\frac{N}{2}\right) \sum_{k}(\alpha
_{k}b_{k}^{\dag }-\alpha _{k}^{\ast }b_{k})-i\Gamma _{\mathrm{loss}},\\
F_{2}(t) &=&\int_{0}^{t}ds\int_{0}^{s}ds^{\prime }[H_{I}(s),H_{I}(s^{\prime
})]  \notag \\
&=&-2iN_{\uparrow }^{2}\sum_{k}\left\vert g_{k}\right\vert
^{2}\int_{0}^{t}ds\int_{0}^{s}ds^{\prime }\sin \omega _{k}(s-s^{\prime
})=-2iN_{\uparrow }^{2}t\Delta (t),
\end{eqnarray}%
with the amplitudes $\alpha _{k}=-ig_{k}\int_{0}^{t}e^{i\omega
_{k}s}ds/t=g_{k}(1-e^{i\omega _{k}s})/\omega _{k}t$,
since $[H_{I}(s),H_{I}(s^{\prime })]=-2iN_{\uparrow }^{2}\sum_{k}\left\vert
g_{k}\right\vert ^{2}\sin \omega _{k}(s-s^{\prime }),$ which commutes with
the high order terms. It is worth to point out that the commutator of the
interaction Hamiltonian at two different times is an operator but not a $C$
number as considering in the single bit case, which can induce the nonlinear
interaction; the noise-induced nonlinear interaction strengthen $\Delta (t)$
can recast as
\begin{eqnarray}
\Delta (t) &=&\frac{1}{t}\sum_{k}\left\vert g_{k}\right\vert
^{2}\int_{0}^{t}ds\int_{0}^{s}ds^{\prime }\sin \omega _{k}(s-s^{\prime })=%
\frac{1}{t}\int_{0}^{\infty }d\omega J(\omega
)\int_{0}^{t}ds\int_{0}^{s}ds^{\prime }\sin \omega (s-s^{\prime })  \notag \\
&=&\frac{1}{t}\int_{0}^{\infty }d\omega J(\omega )\frac{\omega t-\sin
(\omega t)}{\omega ^{2}},
\end{eqnarray}%
where we have used the relation $\sum_{k}\left\vert g_{k}\right\vert
^{2}\rightarrow \int_{0}^{\infty }d\omega J(\omega )$.  Then,
\begin{eqnarray}
U(t) &=&\exp \left[ -iF_{1}(t)-\frac{1}{2}F_{2}(t)\right] =\exp (-i\lambda
tJ_{z})\exp \left[ \left(J_{z}+\frac{N}{2}\right)\sum_{k}(\alpha _{k}b_{k}^{\dag
}-\alpha _{k}^{\ast }b_{k})\right] \exp [itN_{\uparrow }^{2}\Delta
(t)]e^{-t\Gamma _{\mathrm{loss}}}  \notag \\
&=&\exp \left[ -it\lambda ^{\prime }J_{z}\right] \exp \left[ it\Delta
(t)J_{z}^{2}\right] \exp (-t\Gamma _{\mathrm{loss}})\exp [i\phi_0(t)]\exp %
\left[ J_{z}\sum_{k}(\alpha _{k}b_{k}^{\dag }-\alpha _{k}^{\ast }b_{k})%
\right],
\end{eqnarray}%
where $\lambda ^{\prime }=\lambda -N\Delta (t)$ and $\phi_0(t)$ is the global
phase and will be dropped.

\section{ Spin squeezing with $\Gamma_{\mathrm{loss}} \neq 0$}

In this Appendix, we present a detailed of SS with $\Gamma_{\mathrm{loss}}
\neq 0$. To this end, we assume, without loss of generality, that $\hat{n}%
_{0}=(\sin \vartheta \cos \phi ,\sin \vartheta \sin \phi ,\cos
\vartheta )$, where $\vartheta =\tan ^{-1}\left( \sqrt{\langle J_{x}\rangle
^{2}+\langle J_{y}\rangle ^{2}}/\langle J_{z}\rangle \right) $ and $\phi
=\tan ^{-1}\left( \langle J_{y}\rangle /\langle J_{x}\rangle \right) $ are
polar and azimuthal angles, respectively. We then define two mutually
perpendicular unit vectors $\hat{n}_{1}=(-\sin \phi ,\cos \phi ,0)$
and $\hat{n}_{2}=(\cos \vartheta \cos \phi ,\cos \vartheta \sin \phi
,-\sin \vartheta )$. Clearly, both $\hat{n}_{1}$ and $\hat{n}_{2}$ are
perpendicular to $\hat{n}_{0}$ such that $(\hat{n}_{1},\hat{n}_{2},\hat{n}%
_{0})$ form a right-hand frame. Now, the minimal fluctuation of a spin
component perpendicular to the mean spin is

\begin{equation}
(\Delta J_{\hat{n}_{\perp }})_{\mathrm{min}}^{2}=\frac{1}{2}\left[C-\sqrt{%
A^{2}+B^{2}}\right],
\end{equation}%
and the mean spin is
\begin{equation}
|\langle {\mathbf{J}}\rangle |=\sqrt{%
\left\langle J_{x}\right\rangle ^{2}+\left\langle J_{x}\right\rangle
^{2}+\left\langle J_{z}\right\rangle ^{2}}
=\sqrt{%
\left|\langle J_{+}\right\rangle |^{2}+\left\langle J_{z}\right\rangle ^{2}},
\end{equation}
where
\begin{eqnarray}
A &=&\frac{\sin ^{2}\vartheta }{2}\left[ j(j+1)-3\langle J_{z}^{2}\rangle %
\right] \frac{(1+\cos ^{2}\vartheta )}{2}{\mathrm{Re}}[\langle
J_{+}^{2}\rangle e^{-2i\phi }]+\sin \vartheta \cos \theta {\mathrm{Re}}%
[\langle J_{+}(2J_{z}+1)\rangle e^{-i\phi }],  \notag \\
B &=&-\cos \vartheta {\mathrm{Im}}[\langle J_{+}^{2}\rangle e^{-2i\phi
}]+\sin \vartheta {\mathrm{Im}}[\left\langle J_{+}(2J_{z}+1)\right\rangle
e^{-i\phi }],  \notag \\
C &=&j(j+1)-\langle J_{z}^{2}\rangle -{\mathrm{Re}}[\langle J_{+}^{2}\rangle
e^{-2i\phi }]-\frac{\sin ^{2}\vartheta }{2}[j(j+1)-3\langle
J_{z}^{2}\rangle ]+\frac{(1+\cos ^{2}\vartheta )}{2}{\mathrm{Re}}[\langle
J_{+}^{2}\rangle e^{-2i\phi }]-\frac{\sin (2\vartheta )}{2}{\mathrm{Re}}%
[\langle J_{+}(2J_{z}+1)\rangle e^{-i\phi }],  \notag \\
&&
\end{eqnarray}%
with
\begin{eqnarray*}
\left\langle J_{+}\right\rangle &=&je^{it\lambda ^{\prime }}e^{-N\Gamma _{%
\mathrm{loss}}t}e^{-t\gamma (t)}\left\{ \cos [t\Delta (t)]\cosh (\Gamma _{%
\mathrm{loss}}t)+i\sin [t\Delta (t)]\sinh (\Gamma _{\mathrm{loss}}t)\right\}
^{2j-1} \\
\left\langle J_{+}^{2}\right\rangle &=&j\left( j-\frac{1}{2}\right)
e^{-N\Gamma _{\mathrm{loss}}t}e^{-4t\gamma (t)}e^{2it\lambda ^{\prime
}}\{\cos [2(t\Delta (t))]\cosh (\Gamma t)+i\sin [2(t\Delta (t))]\sinh
(\Gamma _{\mathrm{loss}}t)\}^{2j-2}, \\
\left\langle J_{z}\right\rangle &=&\frac{-j}{2^{j}}e^{-N\Gamma _{\mathrm{loss%
}}t}\sinh (2\Gamma _{\mathrm{loss}}t)[1+\cosh (2\Gamma _{\mathrm{loss}%
}t)]^{j-1} \\
\left\langle J_{z}^{2}\right\rangle &=&\frac{je^{-N\Gamma _{\mathrm{loss}}t}%
}{2^{j}(1+e^{2\Gamma _{\mathrm{loss}}t})^{2}}[1+\cosh (2\Gamma _{\mathrm{loss%
}}t)]^{j}[2e^{2\Gamma _{\mathrm{loss}}t}+j(1-e^{2\Gamma _{\mathrm{loss}%
}t})^{2}], \\
\left\langle J_{+}(2J_{z}+1)\right\rangle &=&2j\left( j-\frac{1}{2}\right)
e^{it\lambda ^{\prime }}e^{-N\Gamma _{\mathrm{loss}}t}e^{-t\gamma
(t)}\left\{ \cos [t\Delta (t)]\cosh (\Gamma _{\mathrm{loss}}t)+i\sin
[t\Delta (t)]\sinh (\Gamma _{\mathrm{loss}}t)\right\} ^{2j-2} \\
&&\times \left\{ -\cos [t\Delta (t)]\sinh (\Gamma _{\mathrm{loss}}t)+i\sin
[t\Delta (t)]\cosh (\Gamma _{\mathrm{loss}}t)\right\}.
\end{eqnarray*}

\section{ Matrix elements  of $C_{\perp}$ for $N=2$}

The matrix elements for the symmetric matrix $\boldsymbol{C}$ \ in $yz$
plane are

\begin{eqnarray}
C_{yy} &=&4\left[ \frac{\left\vert \beta _{+}\right\vert ^{2}(p-p_{-})^{2}}{%
p+p_{-}}+\frac{\left\vert \beta _{-}\right\vert ^{2}(p-p_{+})^{2}}{p+p_{+}}%
\right],   \notag \\
C_{zz} &=&4\left[ \frac{\left\vert \alpha _{+}\right\vert ^{2}(p-p_{-})^{2}}{%
p+p_{-}}+\frac{\left\vert \alpha _{-}\right\vert ^{2}(p-p_{+})^{2}}{p+p_{+}}%
\right],   \notag \\
C_{yz} &=&4\sqrt{2}\left[ \frac{(p-p_{-})^{2}\alpha _{+}\mathrm{Im}\beta _{+}%
}{p+p_{-}}+\frac{(p-p_{+})^{2}\alpha _{-}\mathrm{Im}\beta _{-}}{p+p_{+}}%
\right],
\end{eqnarray}%
with
\begin{eqnarray}
p &=&\frac{1}{4}(1-e^{-4t\gamma (t)}),\hspace{0.5cm} p_{\pm }=\frac{1}{8}e^{-4t\gamma (t)}%
\left[ 1+3e^{4t\gamma (t)}\pm \Xi \right],   \notag\\
\Xi &=&\sqrt{(1-e^{4t\gamma (t)})^{2}+16e^{6t\gamma (t)}},
\end{eqnarray}%
\begin{eqnarray}
\alpha _{\pm } &=&\frac{2\sqrt{2}}{\sqrt{16+e^{-6t\gamma (t)}\left[
1-e^{4t\gamma (t)}\pm \Xi \right] ^{2}}}, \hspace{0.5cm} \beta _{\pm }=\frac{-e^{it\Delta
(t)}e^{-3t\gamma (t)}\left[ 1-e^{4t\gamma (t)}\pm \Xi \right] }{\sqrt{%
16+e^{-6t\gamma (t)}\left[ 1-e^{-4t\gamma (t)}\pm \Xi \right] ^{2}}}.
\end{eqnarray}%
\end{widetext}

\end{document}